\documentclass[runningheads]{llncs}

\usepackage[margin=0.9in]{geometry}

\usepackage{algorithmicx}
\usepackage[noend]{algpseudocode}
\usepackage{algorithm}
\algnewcommand{\algorithmicswitch}{\textbf{switch}}
\algdef{SE}[SWITCH]{Switch}{EndSwitch}[1]{\algorithmicswitch\ #1\ \algorithmicdo}{\algorithmicend\ \algorithmicswitch}%
\algtext*{EndSwitch}%

\algnewcommand{\algorithmiccase}{\textbf{case}}
\algdef{SE}[CASE]{Case}{EndCase}[1]{\algorithmiccase\ #1}{\algorithmicend\ \algorithmiccase}%
\algtext*{EndCase}%

\algnewcommand{\algorithmicon}{\textbf{on}}
\algdef{SE}[ON]{On}{EndOn}[1]{\algorithmicon\ #1\ \algorithmicdo}{\algorithmicend\ \algorithmicon}%
\algtext*{EndOn}%

\algrenewcommand{\algorithmicdo}{}
\algrenewcommand{\algorithmicthen}{}

\algnewcommand{\algorithmicgoto}{\textbf{goto}}%
\algnewcommand{\Goto}[1]{\algorithmicgoto~\ref{#1}}%

\algnewcommand{\algorithmicbreak}{\textbf{break}}%
\algnewcommand{\Break}[0]{\algorithmicbreak}%

\algnewcommand{\algorithmicwaiton}{\textbf{wait on}}%
\algnewcommand{\WaitOn}[1]{\algorithmicwaiton~{#1}}%

\PassOptionsToPackage{hyphens}{url}
\usepackage{url}
\usepackage{hyperref}
\hypersetup{breaklinks=true}
\usepackage[utf8]{inputenc}
\usepackage[T1]{fontenc}

\usepackage{amssymb,amsfonts,amsmath,amsthm}
\usepackage{graphicx}
\usepackage{xcolor}

\usepackage{subcaption}
\usepackage{float}

\usepackage{IEEEtrantools}
\allowdisplaybreaks

\usepackage[shortlabels,inline]{enumitem}

\usepackage{xstring}

\usepackage{datetime2}

\usepackage{siunitx}

\usepackage{varwidth}

\usepackage{tikz}
\usetikzlibrary{calc}
\usetikzlibrary{arrows}
\usetikzlibrary{arrows.meta}
\usetikzlibrary{patterns}
\usetikzlibrary{positioning}
\usetikzlibrary{decorations.pathreplacing}
\usetikzlibrary{shapes.misc}
\usetikzlibrary{spy}

\pgfdeclarelayer{bg1}
\pgfdeclarelayer{bg2}
\pgfsetlayers{bg1,bg2,main}

\usepackage{pgfplots}
\pgfplotsset{compat=1.14}
\usepgfplotslibrary{fillbetween}

\usepackage{xspace}
\usepackage{ifthen}

\newcommand{\LOGda}[2]{%
    \ifthenelse{\equal{#1}{}}{%
        \ensuremath{\mathsf{LOG}_{\mathrm{da}}^{#2}}%
    }{%
        \ensuremath{\mathsf{LOG}_{\mathrm{da},#1}^{#2}}%
    }%
}
\newcommand{\LOGbft}[2]{%
    \ifthenelse{\equal{#1}{}}{%
        \ensuremath{\mathsf{LOG}_{\mathrm{bft}}^{#2}}%
    }{%
        \ensuremath{\mathsf{LOG}_{\mathrm{bft},#1}^{#2}}%
    }%
}

\theoremstyle{plain}
\newtheorem*{theorem*}{Theorem}

\definecolor{myParula01Blue}{RGB}{0,114,189}
\definecolor{myParula02Orange}{RGB}{217,83,25}
\definecolor{myParula03Yellow}{RGB}{237,177,32}
\definecolor{myParula04Purple}{RGB}{126,47,142}
\definecolor{myParula05Green}{RGB}{119,172,48}
\definecolor{myParula06LightBlue}{RGB}{77,190,238}
\definecolor{myParula07Red}{RGB}{162,20,47}

\tikzset{myparula11/.style={color=myParula01Blue,solid,mark=+,mark options={solid}}}
\tikzset{myparula12/.style={color=myParula01Blue,densely dashed,mark=x,mark options={solid}}}
\tikzset{myparula13/.style={color=myParula01Blue,densely dotted,mark=o,mark options={solid}}}
\tikzset{myparula14/.style={color=myParula01Blue,dashdotted,mark=triangle,mark options={solid}}}
\tikzset{myparula15/.style={color=myParula01Blue,dashdotdotted,mark=square,mark options={solid}}}

\tikzset{myparula21/.style={color=myParula02Orange,solid,mark=+,mark options={solid}}}
\tikzset{myparula22/.style={color=myParula02Orange,densely dashed,mark=x,mark options={solid}}}
\tikzset{myparula23/.style={color=myParula02Orange,densely dotted,mark=o,mark options={solid}}}
\tikzset{myparula24/.style={color=myParula02Orange,dashdotted,mark=triangle,mark options={solid}}}
\tikzset{myparula25/.style={color=myParula02Orange,dashdotdotted,mark=square,mark options={solid}}}

\tikzset{myparula31/.style={color=myParula03Yellow,solid,mark=+,mark options={solid}}}
\tikzset{myparula32/.style={color=myParula03Yellow,densely dashed,mark=x,mark options={solid}}}
\tikzset{myparula33/.style={color=myParula03Yellow,densely dotted,mark=o,mark options={solid}}}
\tikzset{myparula34/.style={color=myParula03Yellow,dashdotted,mark=triangle,mark options={solid}}}
\tikzset{myparula35/.style={color=myParula03Yellow,dashdotdotted,mark=square,mark options={solid}}}

\tikzset{myparula41/.style={color=myParula04Purple,solid,mark=+,mark options={solid}}}
\tikzset{myparula42/.style={color=myParula04Purple,densely dashed,mark=x,mark options={solid}}}
\tikzset{myparula43/.style={color=myParula04Purple,densely dotted,mark=o,mark options={solid}}}
\tikzset{myparula44/.style={color=myParula04Purple,dashdotted,mark=triangle,mark options={solid}}}
\tikzset{myparula45/.style={color=myParula04Purple,dashdotdotted,mark=square,mark options={solid}}}

\tikzset{myparula51/.style={color=myParula05Green,solid,mark=+,mark options={solid}}}
\tikzset{myparula52/.style={color=myParula05Green,densely dashed,mark=x,mark options={solid}}}
\tikzset{myparula53/.style={color=myParula05Green,densely dotted,mark=o,mark options={solid}}}
\tikzset{myparula54/.style={color=myParula05Green,dashdotted,mark=triangle,mark options={solid}}}
\tikzset{myparula55/.style={color=myParula05Green,dashdotdotted,mark=square,mark options={solid}}}

\tikzset{myparula61/.style={color=myParula06LightBlue,solid,mark=+,mark options={solid}}}
\tikzset{myparula62/.style={color=myParula06LightBlue,densely dashed,mark=x,mark options={solid}}}
\tikzset{myparula63/.style={color=myParula06LightBlue,densely dotted,mark=o,mark options={solid}}}
\tikzset{myparula64/.style={color=myParula06LightBlue,dashdotted,mark=triangle,mark options={solid}}}
\tikzset{myparula65/.style={color=myParula06LightBlue,dashdotdotted,mark=square,mark options={solid}}}

\tikzset{myparula71/.style={color=myParula07Red,solid,mark=+,mark options={solid}}}
\tikzset{myparula72/.style={color=myParula07Red,densely dashed,mark=x,mark options={solid}}}
\tikzset{myparula73/.style={color=myParula07Red,densely dotted,mark=o,mark options={solid}}}
\tikzset{myparula74/.style={color=myParula07Red,dashdotted,mark=triangle,mark options={solid}}}
\tikzset{myparula75/.style={color=myParula07Red,dashdotdotted,mark=square,mark options={solid}}}

\pgfplotsset{
    mysimpleplot/.style = {
        every axis plot/.prefix style={thick},
        width=1.0\linewidth,
        height=0.75\linewidth,
        title style={font=\footnotesize,align=center},
        legend cell align=left,
        legend style={font=\footnotesize},
        legend columns=3,
        legend style={
            at={(0.5,1)},
            yshift=0.3em,
            anchor=south,
            draw=none,
            /tikz/every even column/.append style={
                column sep=0.3em
            },
            cells={
                align=left
            }
        },
        grid=both,
        minor tick num=3,
        major grid style={solid,draw=gray!50},
        minor grid style={densely dotted,draw=gray!50},
        label style={font=\footnotesize,align=center},
        tick label style={font=\footnotesize},
    },
}

\usepackage{comment}
\usepackage{xcolor}
\hypersetup{
    colorlinks,
    linkcolor={red!50!black},
    citecolor={blue!50!black},
    urlcolor={blue!80!black}
}
\usepackage{hyperref}

\begin{document}
\title{Time is Money: Strategic Timing Games \\ in Proof-of-Stake Protocols}
\author{
Caspar Schwarz-Schilling\inst{1} \and
Fahad Saleh \inst{2}\and
Thomas Thiery \inst{1}\and \\
Jennifer Pan \inst{3}\and
Nihar Shah \inst{3}\and
Barnabé Monnot\inst{1}
}

% \authorrunning{}

% If authors are in (no) particular order (un-)comment next line. 
% \let\thefootnote\relax\footnotetext{Authors are listed in no particular order.}


\institute{
Ethereum Foundation\\
\email{\{caspar.schwarz-schilling,thomas.thiery,barnabe.monnot\}@ethereum.org} \and 
Wake Forest University\\
\email{saleh@wfu.edu} \and 
Jump Crypto\\
\email{\{jpan,nshah\}@jumptrading.com}
}
\maketitle % typeset the header of the contribution

\begin{abstract}
We propose a model suggesting that honest-but-rational consensus participants may play \emph{timing games}, and strategically delay their block proposal to optimize MEV capture, while still ensuring the proposal's timely inclusion in the canonical chain. In this context, ensuring economic fairness among consensus participants is critical to preserving decentralization. We contend that a model grounded in honest-but-rational consensus participation provides a more accurate portrayal of behavior in economically incentivized systems such as blockchain protocols.  We empirically investigate timing games on the Ethereum network and demonstrate that while timing games are worth playing, they are not currently being exploited by consensus participants. By quantifying the marginal value of time, we uncover strong evidence pointing towards their future potential, despite the limited exploitation of MEV capture observed at present.

\end{abstract}

\section{Introduction}
\label{sec:introduction}
Consensus protocols are typically evaluated based on their ability to maintain liveness and safety \cite{cachin2017blockchain}, referring to the regular addition of new transactions to the output ledger in a timely manner, and to the security of confirmed transactions remaining in their positions within the ledger. However, beyond liveness and safety, blockchain protocols require fairness of economic outcomes amongst consensus participants to preserve decentralization. More specifically, a protocol should be designed to maximize profitability of honest participation, wherein participants adhere to the prescribed rules. Otherwise, a deviating participant will outcompete their honest peers, leading to centralization of the validation set over time and security implications for consensus itself.

However, the advent of Maximal Extractable Value (MEV) frustrates such fairness goals. It is defined as the value that consensus participants, in their duties as block producers, accrue by selectively including, excluding and ordering user transactions \cite{daian2019flashboys,babel2021clockwork}, MEV has equally substantial implications for the security of consensus protocols. For a system in which transaction fee rewards are dominant, consensus may become unstable due to increased variance in miner rewards \cite{carlsten2016instability}. Similarly, it was argued that a rational actor issuing a  \emph{whale transaction} with an abnormally large transaction fee can convince peers to fork the current chain, further destabilizing consensus \cite{liao2017incentivizing}.
More broadly, understanding and mitigating the impact of MEV on the security and fairness of blockchain networks has become a central concern of protocol designers \cite{huang2021mev}.

As the \emph{whale transaction} highlights, potential MEV accrues over time as users submit transactions and the value of the set of pending transactions increases for the block producer. As a consequence, time is valuable to consensus participants, a feature obviated by the assumption of honest behavior in previous models of consensus. However, we argue that protocols who wish to preserve properties such as economic fairness amongst consensus participants must assume some share of honest-but-rational consensus participation. In particular, the effects of MEV on the consensus participants' incentives must be better understood.

In this paper, we investigate the possibility for block proposers to delay their block proposal as long as possible while ensuring they become part of the canonical chain, aiming to maximize MEV extraction. The reader may note that in Proof-of-Work (PoW)-based leader selection protocols, delaying a proposal bears the risk of losing to a competing block proposer. PoW protocols exhibit an inherent racing condition that prevents these types of strategic delay deviations, or at least make them unprofitable in expectation. Thus, we investigate the implications of MEV on the incentives of consensus participants, particularly block proposers, in a Proof-of-Stake (PoS) context. More specifically, we consider propose-vote type of PoS protocols, where in each consensus round, one leader proposes a block, and a committee of consensus participants is selected in-protocol to vote on the acceptance of that block. This effectively grants block proposers a short-lived monopoly as the only valid proposer for some given round. During this time interval they can attempt to strategically deviate from their assigned block proposal time and delay the release of their block as long as possible in order to extract more MEV, while still ensuring that a sufficient share of attesters see the block in time to vote it into the canonical chain. This behavior leads to an environment in which honest validators earn less than their deviating counterparts, resulting in stake centralization and second-order effects for consensus stability.

\paragraph{Related Work}

To the best of our knowledge, \emph{timing games} have not been formally analyzed in previous literature on Proof-of-Stake. Selfish mining \cite{eyal2018majority}, studied in the context of Proof-of-Work, relies on appropriately timing the release of a block, in order to waste computation of honest miners and earn an outsize share of the rewards. Our timing games are also concerned with strategic behavior to capture a larger share of the total available rewards to consensus participants, yet do not feature the same dynamics as selfish mining in Proof-of-Work, since participants in many PoS-based consensus mechanisms are given a fixed time interval in which to perform their duties.

The security of PoS-based mechanisms has been discussed in terms of chain growth \cite{dembo2020everything} or focusing on the safety and liveness properties of hybrid protocols such as Gasper \cite{buterin2020combining,neu2021ebb}. The economics literature has also examined Proof-of-Stake security with respect to particular attacks such as the double-spending attack \cite{SalehRFS} and 51\% attacks \cite{JRS}. Separately, incentive considerations in the presence of MEV led to the discovery of severe attacks on the Gasper consensus \cite{schwarz-schilling2022three,neu2022two} and protocol changes to address such attacks \cite{d2022no,d2023recent,d2023simple}.

\paragraph{Our Contributions}
Our work models the value of time to consensus participants and explores the potential emergence of timing games in Proof-of-Stake protocols. By understanding the strategic behavior of consensus participants within this model, we gain insights into how these dynamics affect the robustness of consensus protocols to exogenous incentives, and ultimately fairness.
\begin{itemize}
    \item Despite initial pessimism regarding the existence of equilibria in timing games \cite{monnot2022timing}, we formally show how to sustain equilibrium behavior, where it is individually irrational for proposers to deviate from a schedule enforced by attesters, and reward-sharing is fair among participants (Sections~\ref{sec:model} and~\ref{sec:analysis}).
    \item We then investigate whether such timing games might occur in real-world systems (namely, the Ethereum network), using a large, granular data set recording the MEV offered to block proposers over time. We show incidental deviations from the honest protocol specification, highlighting the feasibility of timing games, yet we do not conclude on the existence of intentional timing games (Section~\ref{sec:empirical}).
\end{itemize}

\section{Model}
\label{sec:model}
We model an infinite horizon game among \textit{block proposers} and \textit{attesters}. Time is partitioned into slots $n \in \mathbb{N}$, each of time length $\Delta > 0$. Each slot $n$ has a block proposer $n$ and a unit measure of attesters $A_n = \{ A_{(i,n)} \}_{i \in [0,1]}$ where $A_{(i,n)}$ refers to the $i$th attester within slot $n$.\footnote{Note that we have a continuum of attesters, rather than a discrete set. In Ethereum PoS, over 18,000 attesters emit a vote per slot (as of 2023-05-12).}\par
~\\
The game evolves as follows:
\begin{itemize}

    \item At the beginning of slot $n$, proposer $n$ acts by deciding whether to build on top of the block of proposer $n-1$ and also when to release their own block. More formally, proposer $n$ selects $\phi_{n} \in \{0,1\}$ and $t_{n} \geq n \cdot \Delta$ where $\phi_{n} = 1$ ($\phi_{n} = 0$) refers to proposer $n$ (not) building on top of the block of proposer $n-1$ and $t_{n}$ denotes the time at which proposer $n$ releases their own block. Note that we specify that proposer $n$ cannot release their block before the start of slot $n$ (i.e., $t_{n} \geq n \cdot \Delta$) but that they release their block after the end of the slot. 
    \item After proposer $n$ acts, all slot $n$ attesters act simultaneously. In particular, attester $A_{(i,n)}$ decides whether to attest to the block of proposer $n$ and also the time to release their attestation. More formally, attester $A_{(i,n)}$ select $\nu_{(i,n)} \in \{0, 1\}$ and $\tau_{(i,n)} \geq n \cdot \Delta$ where $\nu_{(i,n)} = 1$ ($\nu_{(i,n)} = 0$) refers to attester $A_{(i,n)}$ (not) attesting to proposer $n$'s block and $\tau_{(i,n)}$ refers to the time that they release their attestation. Notably, attester $A_{(i,n)}$ can attest to the block of proposer $n$ only if they receive the block before releasing their attestation. We let $\delta_{n,(i,n)} \sim exp(\theta^{-1})$ refer to the random time required for the block of proposer $n$ to reach attester $A_{(i,n)}$ where $\theta > 0$ denotes the average communication time across the network and we assume that the slot length is at least double the average communication time across the network (i.e., $\Delta \geq 2 \theta$). In turn, the action of attester $A_{(i,n)}$ is constrained by $\nu_{(i,n)} = 1 \implies \tau_{(i,n)} \geq t_{n} + \delta_{n,(i,n)}$. 
\end{itemize}

\subsection{Block proposers}
\label{subsec:blockproposers}

The pay-off for proposer $n$ is given as follows:

\begin{equation}
U^P(t_{n}, \phi_{n}) =
\begin{cases}
\alpha + \mu \cdot (t_{n} - t_{n_{-}})^+ \text{ if } \chi_{n} = 1\\
0 \text{ otherwise}
\end{cases}
\label{eqn:proposerutil}
\end{equation}

where $\alpha, \mu > 0$ are exogenous constants while $t_{n_{-}}$ corresponds to the time of the most recent canonical block before slot $n$ and $\chi_{n} \in \{0, 1\}$ corresponds to whether the block in slot $n$ is canonical on the blockchain. We introduce the conditions for a block to become canonical in our model in the following, and delay until Section~\ref{sec:modeljust} its interpretation with respect to established consensus models.

Note that we assume that the reward of proposer $n$ increases linearly with time relative to the most recent canonical block so long as block $n$ eventually becomes canonical. This assumption reflects that proposer $n$ accrues incremental MEV over time by delaying the release of their block but that they risk being skipped if they delay release for too long. The time of the most recent canonical block, $t_{n_-}$, is endogenous where slot $n_{-}$ refers to the most recent canonical slot and is thus given explicitly as follows:

\begin{equation}
    n_{-} = \max\{ k \in \mathbb{N} : \chi_{k} = 1, \chi_k \leq n - 1\}
\end{equation}
For a block to be canonical, we require both that it receives sufficiently many successful attestations and that the subsequent block producer builds on top of it. More formally, letting $\tilde{A}_n$ denote the successful attestations for block $n$, $\chi_n$ is given explicitly as follows:
\begin{equation}
    \chi_n = \begin{cases}
        1 & \text{if } \phi_{n+1} = 1, \tilde{A}_n \geq \gamma\\
        0 & \text{otherwise}
    \end{cases}
    \label{eqn:finalization}
\end{equation}
where the number of successful attestations for block $n$ is given as the measure of attesters in slot $n$ voting for block $n$:
\begin{equation}
    \tilde{A}_n = | \{ i \in [0,1] : \nu_{(i,n)} = 1\}|
\end{equation}

\subsection{Attesters} \label{subsec:attesters}

Attester $(i,n)$ receives a pay-off if and only if two conditions are met:

\begin{itemize}
    \item \textbf{Correctness:} A vote by attester $(i,n)$ is correct if their vote is consistent with the canonical blockchain. Recall that the vote of attester $(i,n)$ is given by $\nu_{(i,n)}$ and the eventual canonical status of the block is given by $\chi_n$; thus, this condition is equivalent to $\nu_{(i,n)} = \chi_n$.
    \item \textbf{Freshness:} A vote by attester $(i,n)$ is fresh if it was received by proposer $n+1$ soon enough that it could be included in the block in slot $n+1$ and the block in slot $n+1$ is eventually made canonical. We let $\delta_{(i,n),n+1} \sim exp(\theta^{-1})$ denote the random communication time between attester $(i,n)$ and proposer $n+1$, implying that the first part of this condition equates with $\tau_{(i,n)} + \delta_{(i,n),n+1} \leq t_{n+1}$. Moreover, the second part of this condition equates with $\chi_{n+1} = 1$.
\end{itemize}

For exposition, we normalize the pay-off for attester $(i,n)$ to unity, implying that their pay-off function is given explicitly as follows:

\begin{equation}
U^A(\nu_{(i,n)}, \tau_{(i,n)}) =
\begin{cases}
1 &\text{if } \nu_{(i,n)} = \chi_n, \tau_{(i,n)} + \delta_{(i,n),n+1} \leq t_{n+1}, \chi_{n+1} = 1 \\ \\
0 &\text{otherwise}
\end{cases}
\label{eqn:attesterutility}
\end{equation}

\section{Analysis}
\label{sec:analysis}
\subsection{Equilibrium analysis}
\label{sec:eqanalysis}

There exists a multiplicity of Nash equilibria. In particular, attesters can coordinate to implement proposers acting at any particular time $\Delta^\star \in [0, \Delta]$ within the slot. Formally, we have the following result:

\begin{proposition} Multiple Equilibria\\
~\\
For any $\Delta^\star \in [0, \Delta]$, there exists an equilibrium as follows:\\
~\\
Proposer $n$ selects $t_n$ as follows:
\begin{equation}
    t_n = n \cdot \Delta + \Delta^\star
    \label{eqn:proposertn}
\end{equation}

\noindent and selects $\phi_n$ as follows:

\begin{equation}
    \phi_n = \begin{cases}
1 &\text{if } t_{n-1} \leq (n-1)\cdot \Delta + \Delta^\star \\
0 &\text{otherwise}
\end{cases}
\label{eqn:proposerphi}
\end{equation}

\noindent Attester $(i,n)$ selects $\nu_{(i,n)}$ as follows:

\begin{equation}
    \nu_{(i,n)} = \begin{cases}
1 &\text{if \eqref{eqn:proposertn} and \eqref{eqn:proposerphi} hold}\\
0 &\text{otherwise}
\end{cases}
\label{eqn:attestervote}
\end{equation}

\noindent and selects $\tau_{(i,n)}$ as follows:

\begin{equation}
    \tau_{(i,n)} = \begin{cases}
t_n + \delta_{n,(i,n)} &\text{if  \eqref{eqn:proposertn} and \eqref{eqn:proposerphi} hold}\\
n \cdot \Delta &\text{otherwise}
\end{cases}
\label{eqn:attestertime}
\end{equation}
\label{prop:multeq}
\end{proposition}

Proposition \ref{prop:multeq} arises because a proposer receives a zero pay-off unless her block earns sufficiently many attestations. In turn, if attesters coordinate on voting for a proposer's block only if the proposer releases her block at a particular time, then the proposer earns a strictly positive pay-off only if she releases her block at that particular time. Thus, since a proposer prefers a strictly positive pay-off to a zero pay-off, each proposer optimally releases her block at the release time on which attesters coordinate. 

As an aside, we emphasize that the referenced coordination by attesters is equilibrium behavior. In particular, an attester receives a strictly positive pay-off only if her attestation is correct, and her attestation is correct only if it agrees with the majority of attesters in her slot. As a consequence, when all other attesters vote in one direction, each attester optimally votes in that same direction to avoid a zero pay-off.

\begin{proof}~\\
We begin by establishing that \eqref{eqn:attestervote} - \eqref{eqn:attestertime} are optimal responses for any attester $(i,n)$. Formally, we take as given that all attesters other than $(i,n)$ follow the equilibrium actions \eqref{eqn:attestervote} - \eqref{eqn:attestertime} and also that all proposers follow the equilibrium actions \eqref{eqn:proposertn} - \eqref{eqn:proposerphi}; in that context, we demonstrate that \eqref{eqn:attestervote} - \eqref{eqn:attestertime} maximize \eqref{eqn:attesterutility} and thus these are equilibrium actions for each attester $(i,n)$.

If \eqref{eqn:proposertn} and \eqref{eqn:proposerphi} hold, then $\phi_{n} = 1$ follows directly for all $n \in \mathbb{N}$. Moreover, if all attesters other than $(i,n)$ follow \eqref{eqn:attestervote}, then \eqref{eqn:proposertn} and \eqref{eqn:proposerphi} imply $\nu_{(-i,n)} = 1$ which implies $\tilde{A}_n = 1$ for all $n \in \mathbb{N}$. Then, since \eqref{eqn:proposertn} and \eqref{eqn:proposerphi} imply $\phi_{n} = 1$ for all $n \in \mathbb{N}$ and also $\tilde{A}_n = 1 \geq \gamma$, \eqref{eqn:finalization} therefore implies $\chi_n = 1$ for all $n \in \mathbb{N}$. In turn, since $\nu_{(i,n)} \neq \chi_{n}$ implies the lowest possible pay-off in \eqref{eqn:attesterutility}, we have that $\nu_{(i,n)} = \chi_{n} = 1$ whenever \eqref{eqn:proposertn} and \eqref{eqn:proposerphi} holds. If \eqref{eqn:proposertn} and \eqref{eqn:proposerphi} do not hold, then \eqref{eqn:attestervote} implies $\nu_{(-i,n)} = 0$ which implies $\tilde{A}_n = 0$ for all $n \in \mathbb{N}$. Moreover, \eqref{eqn:finalization} implies $\chi_n = 0$ for all $n \in \mathbb{N}$. In turn, since $\nu_{(i,n)} \neq \chi_{n}$ implies the lowest possible pay-off, we have that $\nu_{(i,n)} = \chi_{n} = 0$ whenever the conjunction of \eqref{eqn:proposertn} and \eqref{eqn:proposerphi} do not hold. Thus, $\nu_{(i,n)} = 1$ is an optimal response if \eqref{eqn:proposertn} and \eqref{eqn:proposerphi} and $\nu_{(i,n)} = 0$ is an optimal response otherwise, thereby establishing \eqref{eqn:attestervote} as the equilibrium action for any attester $(i,n)$.

To establish \eqref{eqn:attestertime} as an optimal response for attester $(i,n)$, note that \eqref{eqn:attesterutility} pointwise decreases in $\tau_{i,n}$ and thus it is optimal to set $\tau_{(i,n)}$ as low as possible subject to feasibility. In general, $\tau_{(i,n)} \geq n \cdot \Delta$ but $\nu_{(i,n)} = 1 \implies \tau_{(i,n)} \geq t_n + \delta_{n,(i,n)} = n \cdot \Delta + \Delta^\star + \delta_{n,(i,n)} > n \cdot \Delta$. As such, whenever $\nu_{(i,n)} = 0$, then $\tau_{(i,n)} = n \cdot \Delta$, whereas whenever $\nu_{(i,n)} = 1$, then $\tau_{(i,n)} = t_n + \delta_{n,(i,n)}$. Then, as per our proof of \eqref{eqn:attestervote}, \eqref{eqn:proposertn} and \eqref{eqn:proposerphi} imply $\nu_{(i,n)} = 1$ which implies $\tau_{(i,n)} = t_n + \delta_{n,(i,n)}$, whereas if either \eqref{eqn:proposertn} or \eqref{eqn:proposerphi} does not hold, then $\nu_{(i,n)} = 0$ which implies $\tau_{(i,n)} =n \cdot \Delta$, which thereby establishes \eqref{eqn:attestertime}.

We conclude by demonstrating that \eqref{eqn:proposertn} - \eqref{eqn:proposerphi} are an optimal response for any proposer $n$. More formally, we take as given that all attesters follow the equilibrium actions \eqref{eqn:attestervote} - \eqref{eqn:attestertime} and also that all proposers other than proposer $n$ follow the equilibrium actions \eqref{eqn:proposertn} - \eqref{eqn:proposerphi}; in this context, we establish that \eqref{eqn:proposertn} - \eqref{eqn:proposerphi}  maximize \eqref{eqn:attesterutility} and thus these are equilibrium actions for each proposer $n$.

Due to \eqref{eqn:attestervote}, any deviation in \eqref{eqn:proposertn} or \eqref{eqn:proposerphi} implies $\nu_{(i,n)} = 0$ for all $(i,n)$ which further implies $\tilde{A}_n = 0$. Then, under such a deviation, \eqref{eqn:finalization} implies $\chi_n = 0$ which implies a zero pay-off as per \eqref{eqn:proposerutil}. Finally, since pay-offs are bounded below by zero, not deviating from \eqref{eqn:proposertn} and \eqref{eqn:proposerphi} necessarily produces a higher pay-off than any such deviation and thus \eqref{eqn:proposertn} and \eqref{eqn:proposerphi} are equilibrium actions.
\end{proof}

\subsection{Model justification}
\label{sec:modeljust}

The model presented in Section~\ref{sec:model} is an idealized description of a blockchain consensus mechanism. A sequence of proposers is selected, each of which is given the right to produce a block for the slot they are assigned to in the sequence. Once the block is released, a set of attesters assigned for the current slot gets to vote for the presence or absence of the block.

When the proposer chooses to build on the previous block, they affirm its place in the canonical chain. There is no block tree: either the current proposer recognizes the block produced by the proposer before them as part of the canonical chain ($\phi = 1$), or they recognize that the previous proposer failed to produce a block which is part of the canonical chain ($\phi = 0$). With the assumption of a continuum of attesters, at equilibrium, sufficiently many votes reach the following proposer, allowing them to make the call on whether or not the previous proposer's block is canonical.

This model resembles the Streamlet protocol \cite{chan2020streamlet}. A proposer submits a block for consideration to the rest of the network. If $\gamma = 2/3$ share of attesters vote the block in, the block is notarized. If attesters do not, e.g., because the block is unavailable, the chain height is not increased, but the next slot starts, giving the opportunity to the next block producer to submit a block for consideration. Leaders extend the longest chain of notarized blocks they have seen.

The model also bears resemblance with the proposed (block, slot) fork choice rule of the Ethereum Gasper protocol \cite{blockslot}, specifically the dynamically available chain produced by the protocol, when $\gamma = 1/2$. In this model of the fork choice, attesters submit a vote attesting to the presence or absence of a block at some given slot. The canonicity of a block is however complicated by the LMD-GHOST rule for block weight accumulation. Obtaining more than half of the attesters' vote may then neither be a sufficient nor a necessary condition to be part of the canonical chain.

Generally, we formulate the hypothesis that most Proof-of-Stake-based leader selection protocols will be exposed to timing games. As long as duties are assigned according to an absolute (wall-clock) time schedule, there exists no pressure to complete duties in a timely manner comparable to the random arrival process of leaders in Proof-of-Work. For instance, PBFT-based finalization protocols such as Tendermint \cite{kwon2014tendermint} or HotStuff \cite{yin2019hotstuff} do not perform a view change until some timeout is reached, which a leader may use to time their release appropriately. While a sufficiently decentralized committee of validators is an existing feature of these protocols, our model further highlights its role in enforcing timeliness at equilibrium, as described in Section~\ref{sec:eqanalysis}.

\section{An empirical case study: Ethereum}
\label{sec:empirical}
Following a formal analysis of the coordination game between proposers and attesters, we now investigate the occurrence of such strategic timing games in real-world systems. To this end, we examine Ethereum, an ideal candidate for the empirical analysis of potential timing games, owing to its mature MEV market structure and the availability of accessible, informative data points.

We show that timing games are indeed worth playing. However, we find that proposers do not delay their block release with the intention to capture more MEV. Instead, we find that delays are mostly due to latency in their signing processes. Thus, we can conclude that timing games are rational to engage in, but do not yet occur to their full possible extent.

\subsection{Consensus mechanism}
The Ethereum consensus mechanism is a composite of two protocols: variants of LMD GHOST \cite{ghost} and Casper FFG \cite{casper}, often referred to together as Gasper \cite{buterin2020combining}. In this paper, we focus exclusively on Ethereum's \emph{available chain} that is built roughly following LMD GHOST. This is because timing games only occur on the available chain. Within this protocol, time progresses in 12-second slots \cite{consensus-specs-beaconchain}. For each slot, one consensus participant, referred to as a validator, is selected as the \emph{block proposer}. According to the honest validator specifications \cite{consensus-specs-validator}, which define the rules for honest protocol participation, a block should be released at the beginning of the slot (0 seconds into the slot). Furthermore, the protocol selects a committee of \emph{attesters} from the validator set who vote on what they consider to be the latest canonical block as soon as they hear a valid block for their assigned slot, or 4 seconds into the slot, whichever comes first \cite{consensus-specs-validator}. We refer to this 4-second mark as the \emph{attestation deadline}. This dynamic, in which block proposers must release their block early enough for attesters to receive it via the peer-to-peer network before the attestation deadline, results from the attestation deadline serving as a coordination Schelling point \cite{schelling1980strategy}. It is worth noting that the honest validator specification prescribes block proposers to release their block at the beginning of the slot, while attesters only attest 4 seconds into the slot (unless a valid block is heard prior to the attestation deadline). This opens up room for block proposers to release their block strategically---i.e., as late as possible while ensuring they accumulate a sufficient share of attestations.

\subsection{Block production process}
To assess the potential benefits of timing games for block proposers, it is important to comprehend the value of time and the process by which MEV opportunities are captured in the block proposing process. In Ethereum, the MEV market structure evolved and matured significantly over time, turning the block production process into an intricate interplay between specialized actors \cite{huang2021mev}. This division of labor enables validators to profit from MEV without engaging in the complex process of identifying MEV opportunities themselves. Instead, validators can outsource the task of building a maximally profitable block to an out-out-protocol block auction process known as MEV-Boost \cite{mev-boost}. 

\emph{Searchers} look for MEV opportunities (e.g., arbitrages), and submit bundles of transactions alongside bids to express their order preference to \emph{block builders}.
Block builders, in turn, specialize in packing maximally profitable blocks using searcher bundles and other available user transactions before submitting their blocks with bids to \emph{relays}. Relays act as trust facilitators between block proposers and block builders, validating blocks received by block builders and forwarding only valid headers to validators. This ensures validators cannot steal the content of a block builder's block, but can still commit to proposing this block by signing the respective block header. In the long run, Ethereum's plans include enshrining this currently out-of-protocol mechanism into the protocol \cite{pbs-1,pbs-2} to eliminate relay trust assumptions. It is worth noting that MEV-Boost is an opt-in protocol, and validators can always choose to revert to local block building. Finally, when a validator is selected to propose a block in a given slot, they request the highest-bidding block header from the relay, sign it, and return the signed block header to the relay, which then releases the block to the peer-to-peer network. 

In summary, searchers find MEV opportunities and express their transaction-ordering preferences within a block via bids. Block builders aim to build maximally profitable blocks using searcher bundles and user transactions, then submit their block content and bids to relays. Validators ultimately request the highest-paying block header, sign it and return it to relays, which release the signed block to the peer-to-peer network. Due to competition at all levels in this block production process (except for block proposing monopoly), the block proposer is able to capture most of the MEV via this block auction.

\subsubsection{MEV-Boost block auction}

\begin{figure}[ht]
    \centering
    \includegraphics[width=0.8\textwidth]{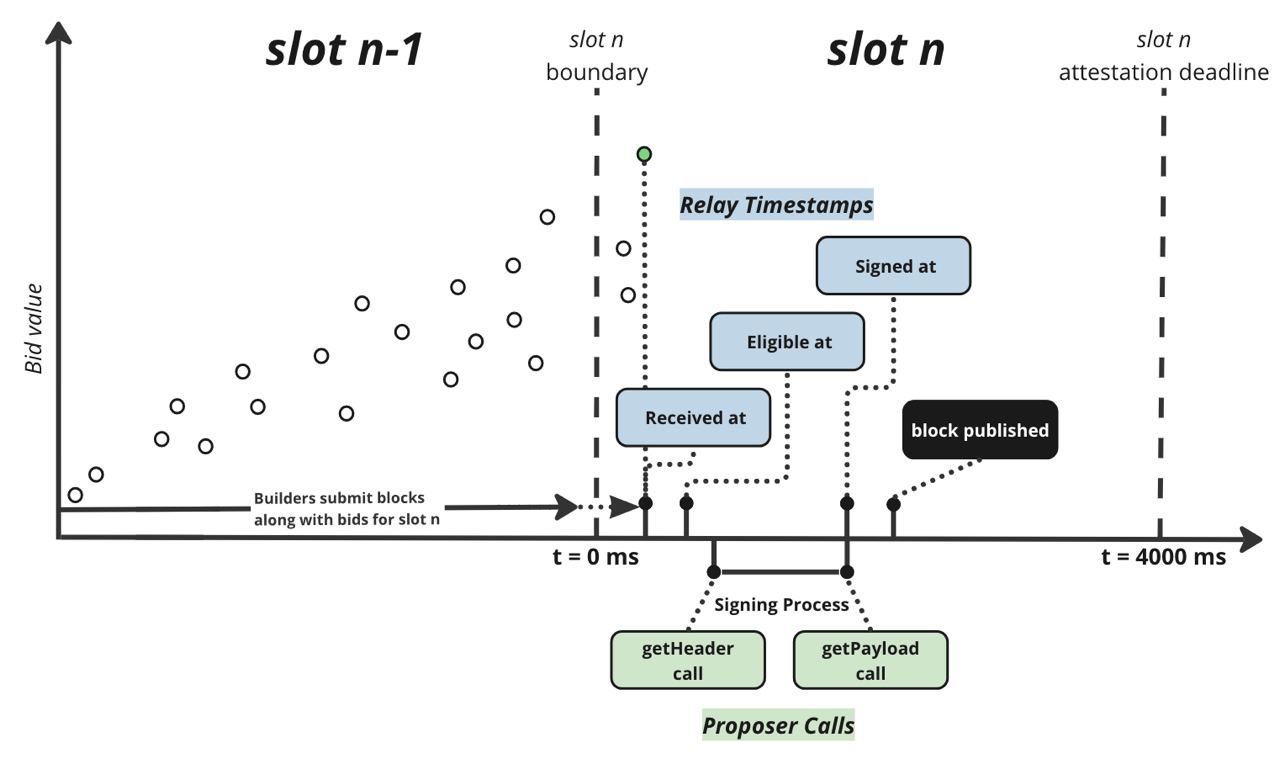}
    \caption{Logical representation of the block production process for slot $n$. Builder bids begin streaming in during slot $n-1$, after which the proposer and relay interact through requests and responses.}
    \label{fig-block-production}
\end{figure}

Here, we granularly outline the sequence of events that take place during the block construction of MEV-Boost block auctions on the Ethereum network. Figure~\ref{fig-block-production} illustrates these events along with their corresponding timestamps, and is intended to serve as a reference for the remainder of this empirical analysis.

The auction for block of slot $n$ begins in slot $n-1$ (at $t=-12000\text{ms}$), during which builders submit blocks alongside bids to relays. This competitive process between block builders determines the right to construct the block for slot $n$ and secures potential MEV-derived profits (block building profit equates to extracted MEV minus bid value). For each bid, the relay logs the timestamps of events at which the bid was received by the relay (\texttt{receivedAt}). After some validity checks are completed by the relay, the bid is made available to the proposer (\texttt{eligibleAt}). When the proposer chooses to propose a block \footnote{An honest participant will request the block header shortly before slot $n$ such that the block can be released on time, at the beginning of slot $n$ ($t=0\text{ms}).$}, the proposer requests \texttt{getHeader} to receive the highest bidding, eligible block header from the relay. Upon receiving the header associated with the winning bid, the proposer signs it and thereby commits to proposing this block built by the respective builder in slot $n$. The signed block header is sent to the relay, along with a request to get the full block content from the relay (\texttt{getPayload}). Finally, the relay receives the signed block header (\texttt{signedAt}) and publishes the full block contents to the peer-to-peer network and proposer. As soon as peers see the new block, validators assigned to the slot can attest to it. This cycle completes one round of consensus repeating every slot.

\subsection{Data sets}
The analysis utilizes data provided by the \emph{ultra sound relay} from March 4, 2023, to April 11, 2023. This covers just under 185,000 slots, interspersed from slot 5,965,398 to slot 6,282,397, and includes all bids placed by block builders through this relay. There were over 800 bids per slot, for a total of over 150 million bids. The winning block originated from the \emph{ultra sound relay} for nearly 85,000 of these slots, and so we measure timestamps and other properties for those slots when investigating winning bids specifically. Finally, we augmented the winning slots with various on-chain measures from the execution layer (EL) and consensus layer (CL), such as attestations and aggregations, using a combination of analytical tools like Dune and direct observation of the peer-to-peer network.

\subsection{Are timing games worth playing?}

\subsubsection{Marginal value of time}
\begin{figure}[ht]
    \centering
    \includegraphics[width=0.8\textwidth]{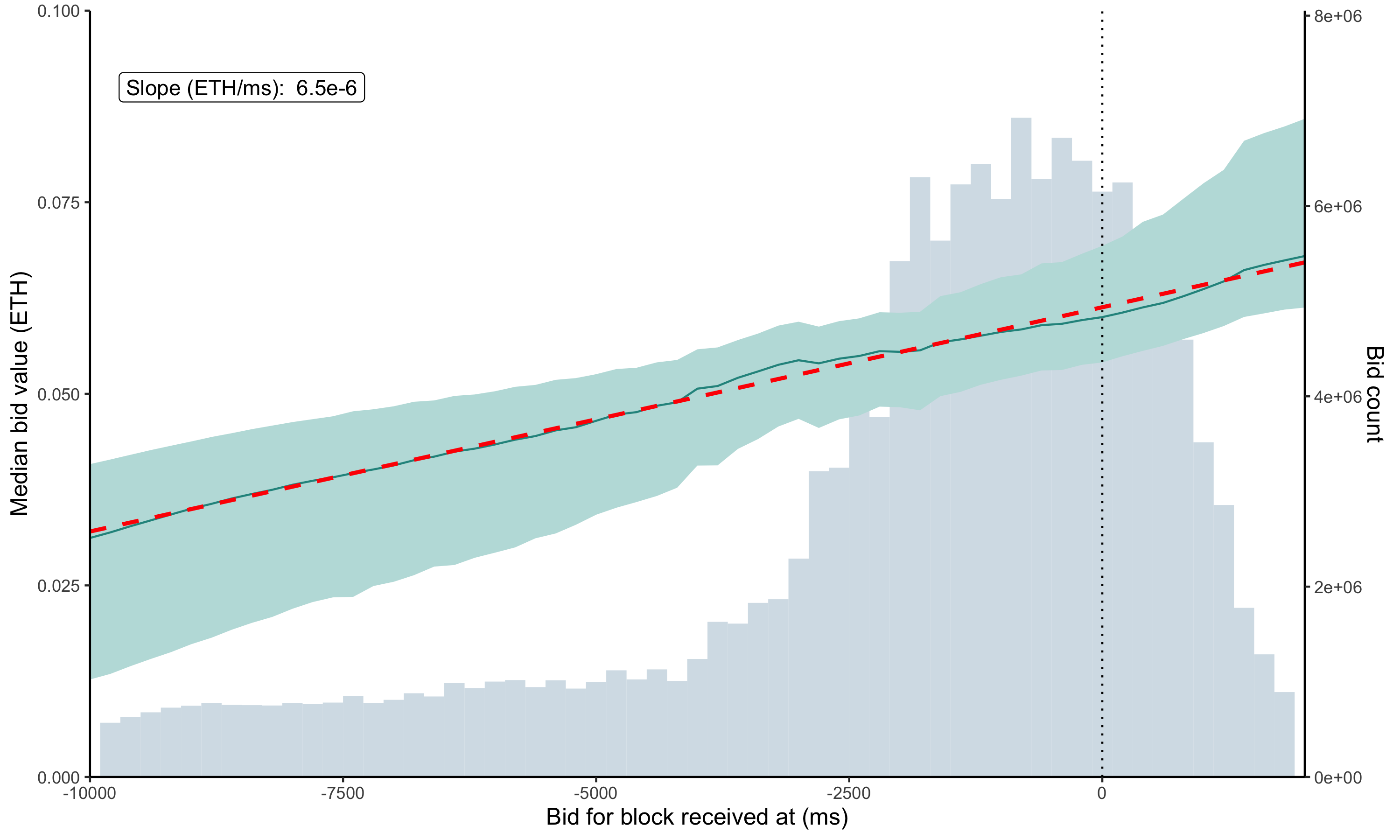}
    \caption{Analysis of bid values and their distribution over slot duration. The histogram (in blue) shows the distribution of bid counts across time in seconds. The dark green line represents the median bid value in Ether (ETH) for each time bin (with its associated IQR in green), residualized against the slot fixed effects that are estimated in a linear regression of bid on timestamp (dashed red line). The x-axis shows time in milliseconds relative to the slot boundary, the left y-axis displays the residualized bid value in ETH, and the right y-axis displays the count of bids.}
    \label{fig-time-value}
\end{figure}

Timing games offer potential for substantial profit due to the increased MEV opportunities they provide. First, we assess whether timing games are worth playing for proposers, by estimating the incremental MEV gained per second. We utilize all bids submitted by builders from the \emph{ultra sound relay} to examine the relationship between the timestamp at which the relay received a bid submitted by a builder (\texttt{receivedAt} timestamp relative to the slot boundary) and the bid value, residualized against slot fixed effects to account for differences between low- and high-MEV regimes and other unobservable forms of heterogeneity. We then fit a regression line to this relationship, obtaining a slope with a coefficient of 0.0065 ETH per second, which represents our estimate for the marginal value of time. Figure~\ref{fig-time-value} depicts the linear increase in median bid values over the slot duration on a point-by-point basis, and the distribution of bid receival times, indicating that most bids are submitted between four seconds before the slot boundary to one second after. This analysis shows there exists a positive marginal value of time, indicating that a rational block proposer would participate in timing games.

\subsection{Are block proposers playing timing games?}
Having shown that timing games are worth playing, we turn our attention to whether proposers are currently taking advantage of the opportunity to accumulate more MEV by committing to a bid later than foreseen by the honest validator specifications.

\subsubsection{Characterizing late block signing behavior}
\begin{figure}[ht]
    \centering
    \begin{subfigure}[b]{0.5\textwidth}
        \centering
        \includegraphics[height=0.4\textheight]{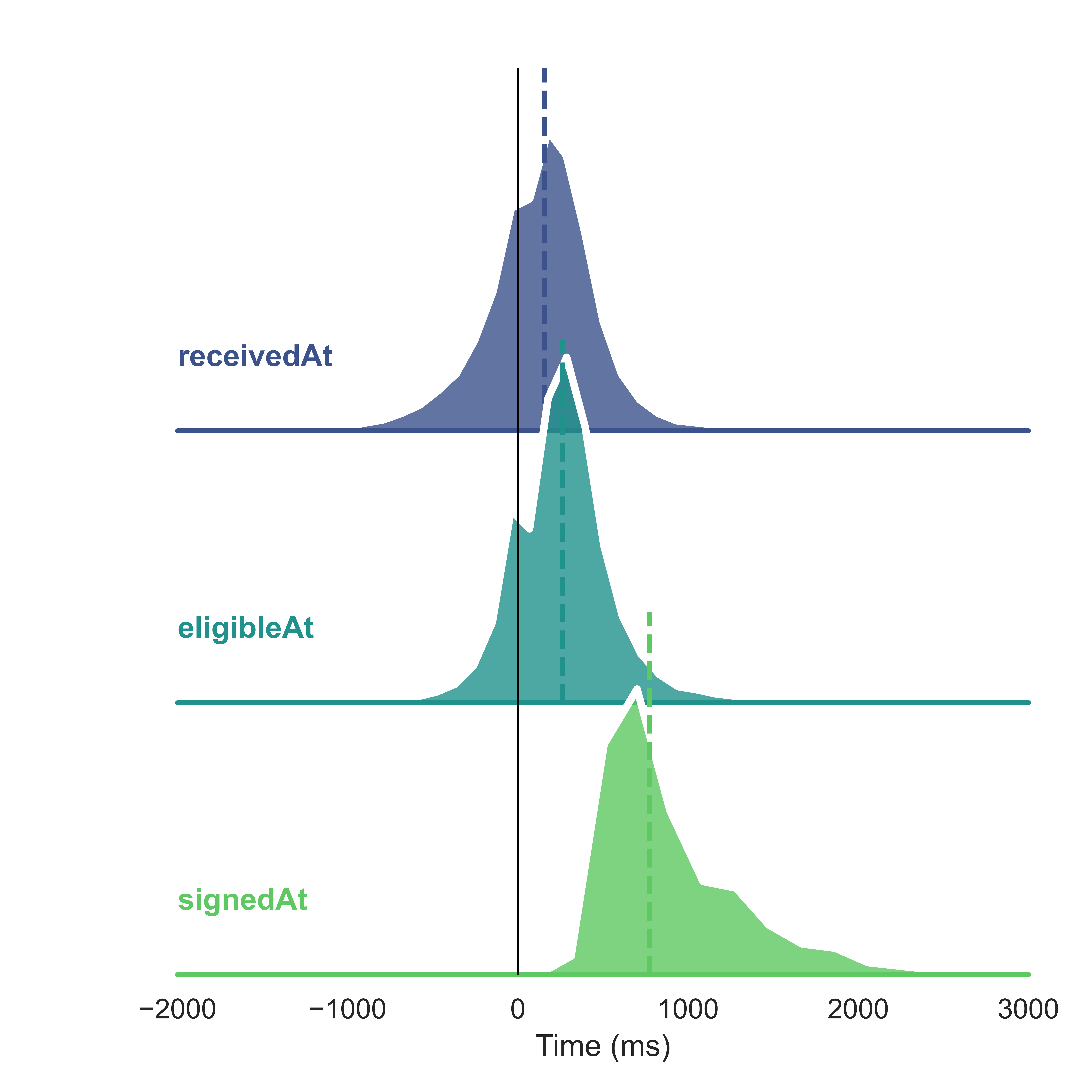}
        \caption{}
        \label{fig-block-timings-a}
    \end{subfigure}%
    \begin{minipage}[b]{0.5\textwidth}
        \begin{subfigure}[b]{\textwidth}
            \centering
            \includegraphics[height=0.2\textheight]{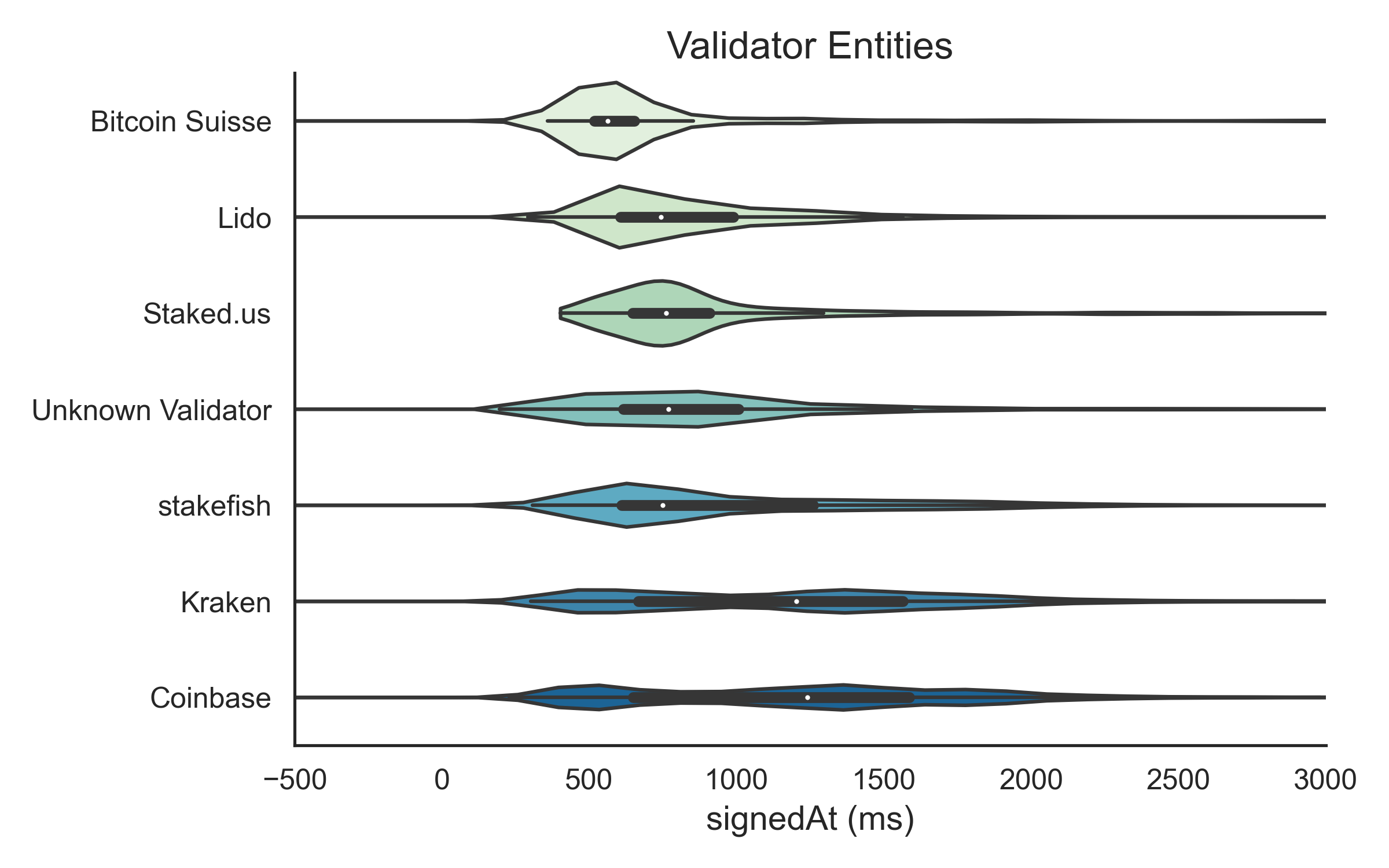}
            \caption{}
            \label{fig-block-timings-b}
        \end{subfigure}
        \begin{subfigure}[b]{\textwidth}
            \centering
            \includegraphics[height=0.2\textheight]{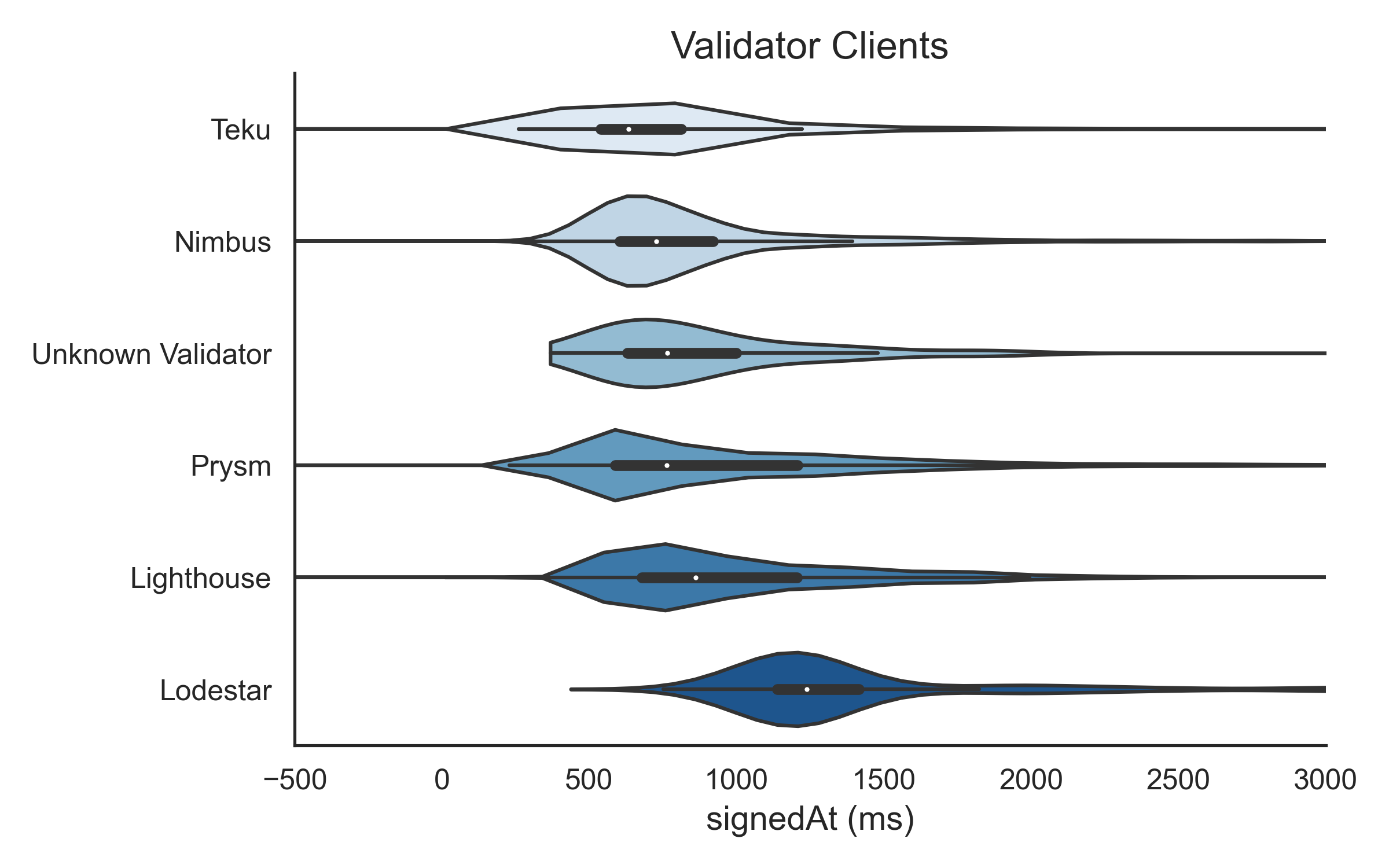}
            \caption{}
            \label{fig-block-timings-c}
        \end{subfigure}
    \end{minipage}
    \caption[]{Analysis of event timestamps and their distributions among Validator Clients and Entities.
    (\ref{fig-block-timings-a}) Multiple Kernel Density Estimation (KDE) distributions of event timestamps from the relay data, showing the probability density functions for three event types: \texttt{receivedAt} (blue), \texttt{eligibleAt} (green), and \texttt{signedAt} (light green). 
    (\ref{fig-block-timings-b}-\ref{fig-block-timings-c}) Violin plots comparing the distribution of \texttt{signedAt} event timestamps for the top 7 validator entities and clients. The x-axis represents time in milliseconds (ms) relative to the slot boundary, while the y-axis displays validator clients and entities, ordered by the mean \texttt{signedAt} time. The width of each violin plot signifies the kernel density estimation of the \texttt{signedAt} event timestamps, demonstrating the distribution and frequency of the events within each group.}
    \label{fig-block-timings}
\end{figure}

First, we investigate whether block headers and associated bids are signed by proposers later than the slot boundary ($t=0$), the time stipulated by the honest protocol specifications to broadcast their block to the network. We observe that winning bids are signed by proposers approximately 774 ms after the slot boundary ($t_{(111573)} = 575.5$, $p < 1 \times 10^{-20}$, using a two-tailed paired Student $t$ test)  and about 513 ms after the relay made the bid eligible  ($t_{(111573)} = 472.6$, $p < 1 \times 10^{-20}$, using a two-tailed paired Student t-test). Figure~\ref{fig-block-timings-a} displays the distribution of timings for winning bids, based on \emph{ultra sound relay} timestamps for bid reception from the builder (\texttt{receivedAt}, median $=157\text{ms}$), eligibility for proposer signing (\texttt{eligibleAt}, median $=260 \text{ms}$), and the actual signing by proposers (\texttt{signedAt}, median $=774\text{ms}$). To better understand the reasons behind late-signing behavior by proposers, we map validator public keys to their staking entities and CL clients, see Figure~\ref{fig-block-timings-b} and~\ref{fig-block-timings-c} respectively). Validator to staking entity mappings were obtained via a combination of open source data sets \footnote{Dune Spellbooks: \url{https://dune.com/spellbook}, Mevboost.pics Open Data: \url{https://mevboost.pics/data.html}}, and validator to client mappings were obtained using blockprint \cite{blockprint}, an open source tool assigning client labels to validators based on their attestation packing on the Ethereum beacon chain. We found that staking entities such as Kraken ($t = 38.9$, $p < 1 \times 10^{-20}$) and Coinbase ($t = 67.6$, $p < 1 \times 10^{-20}$), as well as proposers using the Lodestar client ($t = 44.9$, $p < 1 \times 10^{-20}$) sign block headers significantly later than other block proposer types (results were obtained using two-tailed unpaired Student t-tests). Notably, additional analyses are required to differentiate the interdependencies between validator entities and clients to better understand their roles in late signing behavior. This analysis confirms that proposers are signing blocks significantly later than expected, but it does not yet clarify the underlying reasons, which could include participation in timing games or increased latency for independent reasons, e.g., longer signing processes. 

\begin{figure}[ht]
    \centering
    \begin{subfigure}[b]{0.45\textwidth}
        \centering
        \includegraphics[width=\textwidth]{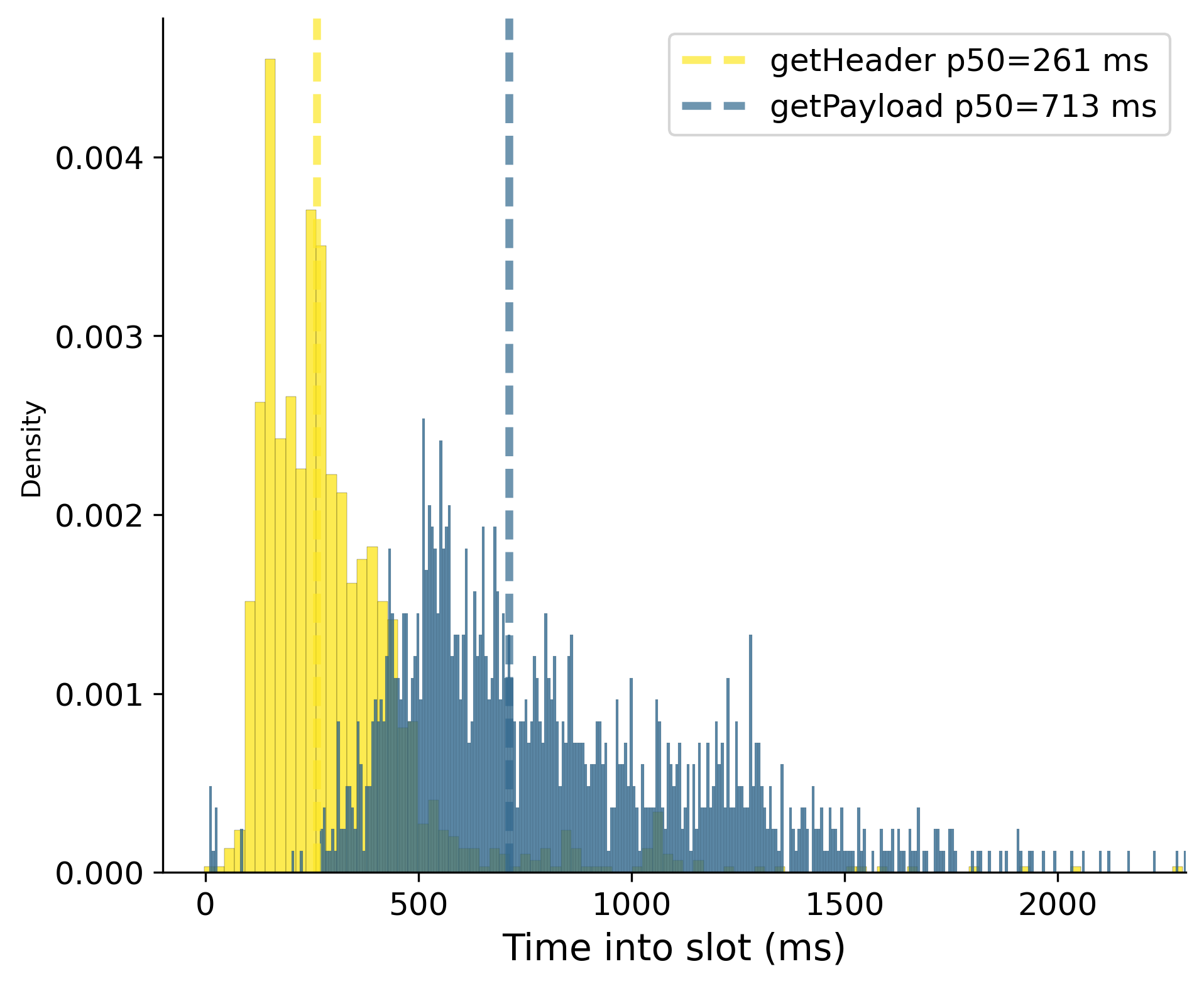}
        \caption{}
        \label{fig:4a}
    \end{subfigure}%
    ~ %add desired spacing between images, e. g. ~, \quad, \qquad etc. 
    \begin{subfigure}[b]{0.45\textwidth}
        \centering
        \includegraphics[width=\textwidth]{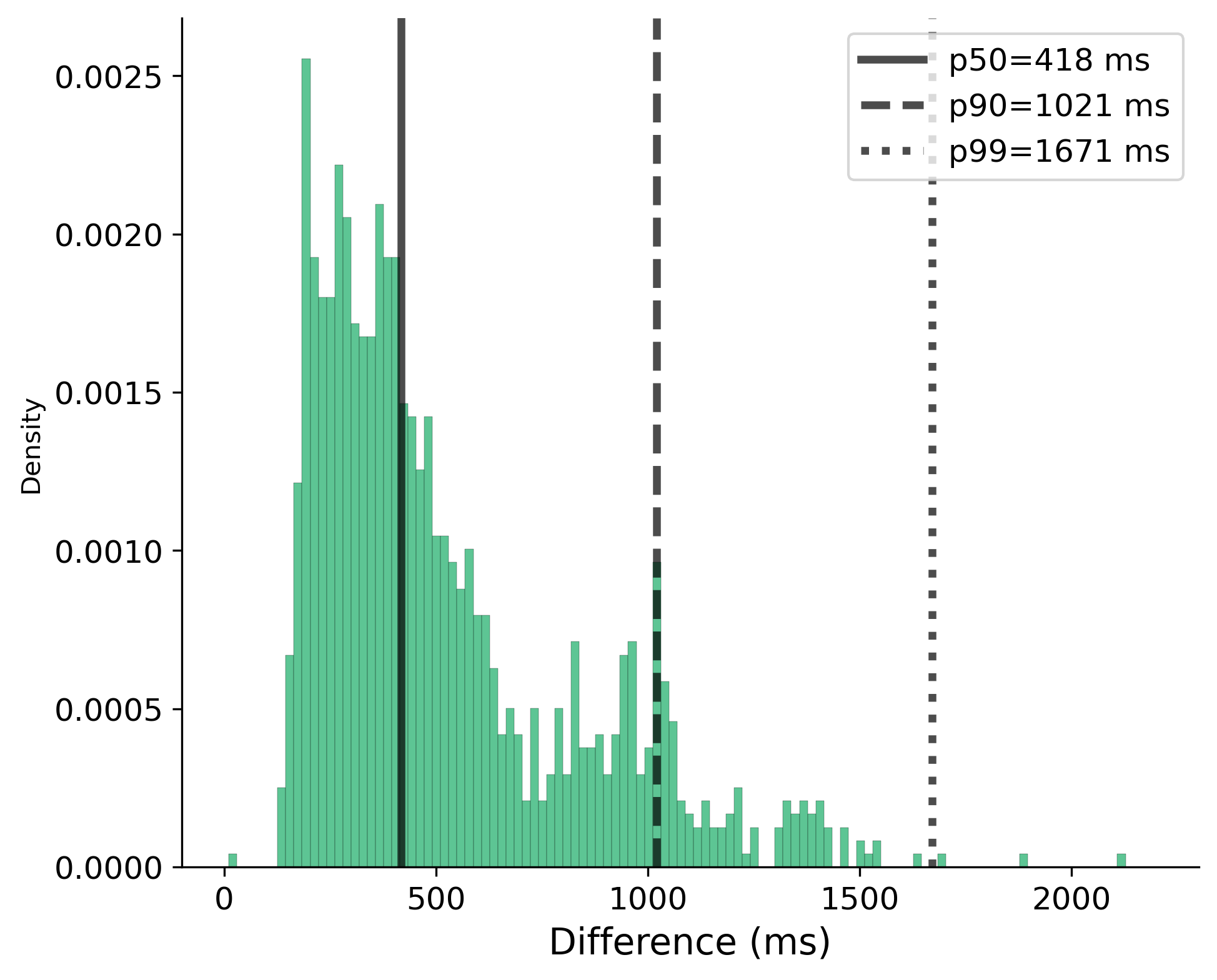}
        \caption{}
        \label{fig:4b}
    \end{subfigure}
    \caption[]{Estimating the latency induced by the signing process.
    (\ref{fig:4a}) Histogram of \texttt{getHeader} and \texttt{getPayload} call timestamps relative to slot boundary. The histogram displays the density of events occurring at different times into the slot (in milliseconds) for \texttt{getHeader} (yellow) and \texttt{getPayload} (blue) calls.    
    (\ref{fig:4b}) Histogram of the time difference between \texttt{getHeader} and \texttt{getPayload} calls. The histogram shows the density of time differences (in milliseconds) \texttt{getHeader} and \texttt{getPayload} calls. Vertical lines represent the $50^\text{th}$ (solid), $90^\text{th}$ (dashed), and $99^\text{th}$ (dotted) percentiles of the distribution.}
    \label{fig-signing-latency}
\end{figure}

We subsequently collected data for 1241 slots (slot 6,200,251 to 6,204,957 on April 11, 2023) and used the time difference between \texttt{getHeader} and \texttt{getPayload} calls by proposers as an approximation to estimate the duration of the signing process (see Figure~\ref{fig-block-production}). Figure~\ref{fig-signing-latency} shows that the median difference between \texttt{getHeader} and \texttt{getPayload} call is 418 ms. Interestingly, this delay, attributable to the signing process, accounts for 75.42\% of the overall latency. This percentage was determined by calculating the difference between the signing time and the moment the bid was deemed eligible by the relay on a slot-by-slot basis, using the formula $\frac{\mathrm{median}(\texttt{getPayload} - \texttt{getHeader})}{\mathrm{median}(\texttt{\texttt{signedAt}} - \texttt{\texttt{eligibleAt}})} \times 100$. We conclude that late signing behavior is primarily attributed to latency caused by the signing process, rather than intentional delays to incorporate more MEV in blocks. This finding aligns with the hypothesis that large US-based staking entities, such as Coinbase and Kraken, may prefer utilizing sophisticated remote secure signing mechanisms, resulting in a lengthier signing process compared to other parties.

\subsection{The impact of latency on the peer-to-peer network}
\begin{figure}[ht]
    \centering
    \begin{subfigure}[b]{0.45\textwidth}
        \centering
        \includegraphics[width=\textwidth]{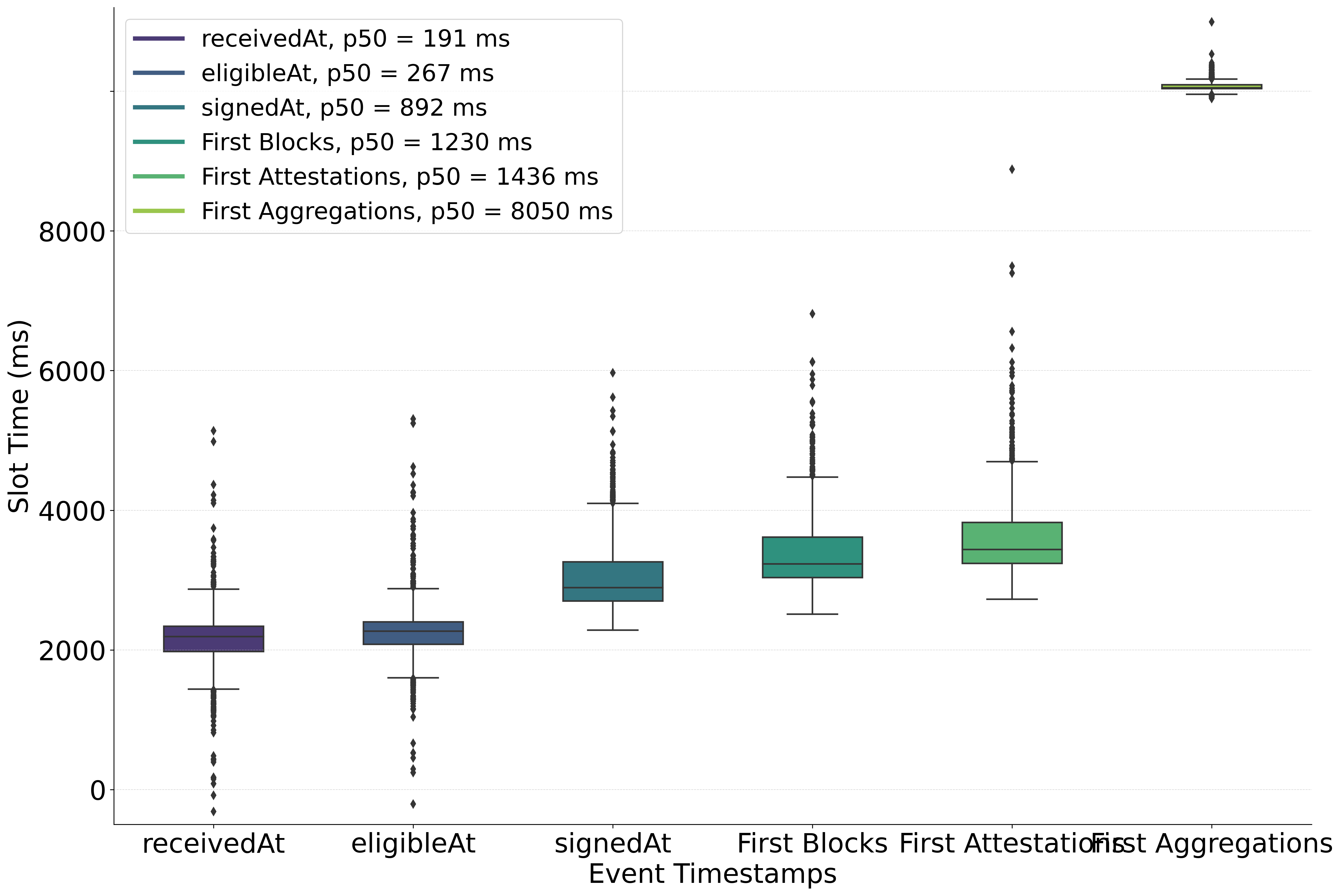}
        \caption{}
        \label{fig-latency-implications-a}
    \end{subfigure}%
    ~ %add desired spacing between images, e. g. ~, \quad, \qquad etc. 
    \begin{subfigure}[b]{0.45\textwidth}
        \centering
        \includegraphics[width=\textwidth]{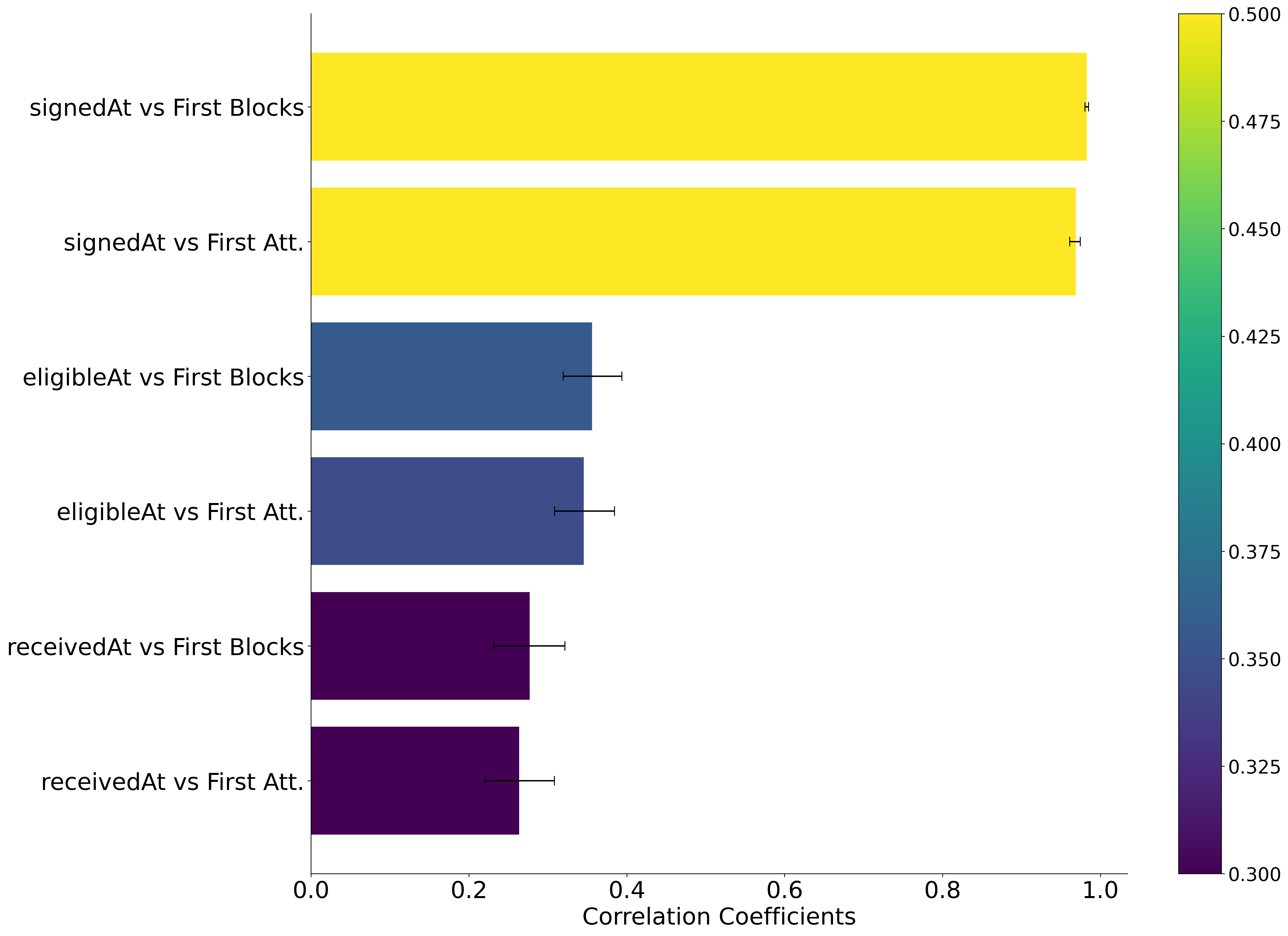}
        \caption{}
        \label{fig-latency-implications-b}
    \end{subfigure}
    \caption[]{Analysis of relay and consensus layer timestamps.
    (\ref{fig-latency-implications-a}) Box plot of the time differences between relay and consensus timestamps. The box plots display the distribution of time differences for \texttt{receivedAt}, \texttt{elligibleAt}, and \texttt{signedAt} events, as well as blocks, attestations, and aggregations first seen by the peer-to-peer network. The boxes represent the interquartile range (IQR) from the first quartile (Q1, $25^\text{th}$ percentile) to the third quartile (Q3, $75^\text{th}$ percentile), while the whiskers extend to the minimum and maximum values within three times the IQR. The horizontal lines within the boxes represent the median values.
    (\ref{fig-latency-implications-b}) Bar plot of Pearson correlation coefficients for each pair of event timestamps. The bars represent the mean correlation coefficient for each relationship, while the error bars represent the 95\% confidence intervals obtained via bootstrapping.
    }
    \label{fig-latency-implications}
\end{figure}

Our prior results indicate that validators are not engaging in timing games to accrue more MEV. Nonetheless, we assess the implications of late signing of consensus messages on the peer-to-peer network. Specifically, we examine the relationship between the relay timestamps, the timings at which blocks are (1) first seen by the rest of the peer-to-peer network and (2) begin collecting attestations and aggregations. The consensus layer data was obtained through nodes run by the Ethereum Foundation, for 2643 slots (slots 6,357,601 to 6,363,807 on May 3 and 4, 2023). Figure~\ref{fig-latency-implications-a} shows the sequence of these event timestamps over the course of a slot. We subsequently assess the correlations between each of these event pairs, as depicted in Figure~\ref{fig-latency-implications-b}. Our analysis reveals high correlations between the time at which blocks are signed by proposers (i.e., the \texttt{signedAt} relay timestamp) and the time at which blocks (correlation coefficient = 0.986) and attestations (correlation coefficient = 0.971) are initially observed by the peer-to-peer network. These findings underscore the significance of proposers signing blocks promptly, as it considerably impacts the downstream processes at the consensus layer in the network.

Next, we evaluate the impact of latency induced by late signing behavior on attestations collected by winning blocks proposed to the peer-to-peer network. We examine the relationship between the time at which blocks are signed by proposers (\texttt{signedAt}), and the share of attestations included by blocks in their respective target slot referred to as slot $n$ in Figure~\ref{fig-block-production}. As a reminder, attestations collected on a given slot $n$ are only included on-chain one (slot $n+1$) or more slots later. In our analysis, we focus on the attestations included in the subsequent slot and compute a metric \emph{next-slot shares}. This metric refers to the percentage of attestations for the winning block in a given slot that appear in the next block (slot $n+1$), out of the total number of attestations in the next slot that refer to any block in the target slot. Our hypothesis is that if a block is signed too late by a proposer, it will not propagate early enough for attesters to vote for it before their attestation deadline ($t=4000\text{ms}$, see Figure~\ref{fig-block-production}, and \cite{consensus-specs-validator}). Hence, in such settings attesters vote for another block (e.g., the parent block), and this will be reflected in the next-slot shares metric.

\begin{figure}[ht]
    \centering
    \includegraphics[width=0.8\textwidth]{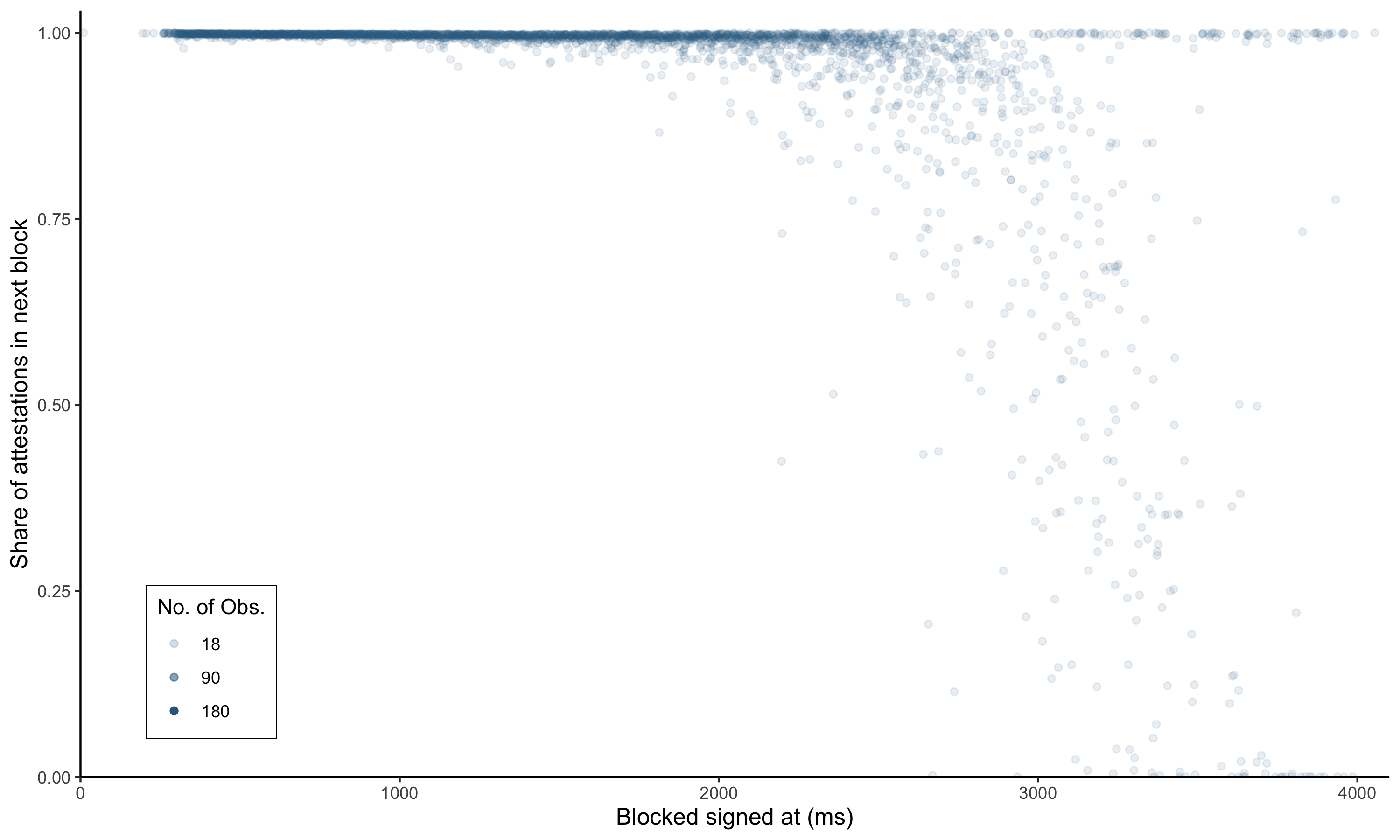}
    \caption{Effects of block signing times on next-slot attestations. This scatter plot features the x-axis displaying the time at which proposers signed the winning block relative to the slot $n$ boundary, and the y-axis illustrating the share of next-slot ($n+1$) attestations for the winning block. Each point on the graph corresponds to the time (in milliseconds) at which the winning block was signed within the slot and the average share of attestations it received, included in the next slot, across all winning blocks signed that specific time.}
    \label{fig-attestation-share-included}
\end{figure}

Figure~\ref{fig-attestation-share-included} shows that latency does indeed cause a steep drop-off in the share of attestations received by the winning block. We observe that the share value stays close to one as long as the block is signed within the first two seconds of the slot. Once the two second threshold is crossed, there is a substantial drop-off and many winning blocks earn fewer than half of the next-slot attestations, which continues to rapidly decrease towards zero as we approach the theoretical $t=4000\text{ms}$ attestation deadline. These results demonstrate the impact of latency on the rest of the peer-to-peer network and highlight the importance of signing and broadcasting blocks on time to prevent missed slots and reorganizations.

We previously documented a private incentive for proposers to delay their block release, according to the steady increase of MEV as time progresses through the slot. Such malicious behavior is not prevalent, as 75\% and 98\% of all blocks are seen by our nodes after two and four seconds into the slot, respectively. Yet, our analysis reveals on-chain evidence whenever latency degrades consensus formation.

\section{Discussion}
\label{sec:discussion}
In this paper, we present an argument that consensus participants are subject to exogenous incentives, primarily MEV, that exist outside the consensus mechanism itself. This highlights the imperative for blockchain protocols to ensure economic fairness among all consensus participants. Specifically, it necessitates a design where honest and honest-but-rational consensus participation become indistinguishable, and honesty within the protocol is the most profitable strategy. This approach ensures that honest-but-rational participants have no incentive to deviate from honest consensus participation.

We present a model that highlights the time-dependent value for consensus participants and probes into the strategic timing considerations that block proposers face. Our model uncovers a spectrum of equilibria wherein attesters can enforce any deadline for block proposals to achieve canonical status, thereby emphasizing the crucial role of Schelling points as coordination mechanisms. For instance, in the Ethereum network we observe the emergence of such Schelling points through the default settings of client software. The widespread use of these default settings among consensus participants generally ensures their effectiveness. 

We support our theoretical findings by observations of the Ethereum network. Our analysis demonstrates that timing games are indeed worth playing for block proposers, enabling them to capture additional MEV by delaying their block proposals beyond the timeframe prescribed by the honest validator specification. However, we observe that current instances of delayed block proposals are primarily due to latency in the block signing process, rather than a conscious strategy to maximize profits. The apparent lack of maximal MEV capture by honest proposers could be attributed to either a lack of common knowledge, existing social norms around this practice. It's clear, however, that these are not sustainable safeguards for maintaining economic fairness.

The implications of timing games are manifold and significant. An honest-but-rational participant who engages in timing games will outperform honest participants, leading to a centralization of stake over time. Hypothetically, this could culminate in a breach of consensus security. In a more practical sense, it may encourage individual stakers to delegate their stake to professional entities adept at these practices, negatively impacting the network's decentralization. Moreover, timing games can overload the messaging system within a short time span, potentially causing cascading failures at the peer-to-peer layer, particularly within client systems.

Essentially, timing games are facilitated by the monopolistic right that block proposers possess for a single round of consensus. Introducing competition in block proposing, similar in effect to the exogenous randomness in Proof of Work (PoW) systems, emerges as a potential solution. However, the challenge lies in deterministically selecting a winning proposer, or reverting to peer-to-peer latency races, which in itself is centralizing. Alternatively, an on-chain heuristic for timely block proposals could incentivize timely participation, yet the allure of MEV rewards might still outweigh any in-protocol consensus rewards. Tackling the root cause of timing games remains an open challenge.

In the Ethereum context, a late-block reorging mechanism has been adopted in the fork choice, effectively imposing a 4-second deadline for block proposers. This constraint significantly limits the extent to which block delays are possible. Looking ahead, the adoption of (block, slot) type of attestations is likely, further refining the protocol. However, it remains challenging to address the root cause of timing games, as it is deeply intertwined with the fundamental workings of Proof of Stake (PoS). Although limiting the length of the proposer's interval is feasible, completely eliminating the monopolistic market structure of block proposers proves to be a difficult task. 

Consequently, it may prove valuable to find a more general abstraction for PoS type of protocols and further explore the implications of consensus participants being exposed to incentives outside of consensus itself, such as MEV. More generally, assuming honest-but-rational as opposed to honest type of consensus participation should prove significant in designing economically fair blockchain protocols.

\section*{Acknowledgments}

The authors acknowledge helpful discussions and comments from Francesco d'Amato and Anders Elowsson. We also appreciate the significant contributions of Mike Neuder in obtaining the necessary data for this study.

% \printbibliography
\bibliographystyle{splncs04}
\bibliography{references}

\appendix

\end{document}